\documentclass[apjl]{emulateapj}

\newcommand{\OIII}{[\mbox{O\,\textsc{iii}}]}
\newcommand{\oiii}{[\mbox{O\,\textsc{iii}}]}

\newcommand{\nii}{[\mbox{N\,\textsc{ii}}]}

\newcommand{\mgb}{\mbox{Mg\,\textsc{i}b}}

\newcommand{\kms}{km s$^{-1}$}
\newcommand{\Ha}{H$\alpha$}     
\newcommand{\ha}{H$\alpha$}   
  
\newcommand{\hb}{H$\beta$}

\newcommand{\ergs}{erg s$^{-1}$} 
\usepackage[T1]{fontenc}
\usepackage{lmodern}
\usepackage [normalem]{ulem}
\usepackage{color}
\usepackage{hyperref}
\usepackage{natbib}
\usepackage{tikz}
\usepackage{graphicx}
\usepackage{bigstrut}
\usepackage[export]{adjustbox}
\usepackage{afterpage}
\usepackage{enumitem}
\setlist[enumerate]{itemsep=0mm}
\usetikzlibrary{positioning,fit,calc}
\usetikzlibrary{shapes,backgrounds}
\usepackage{float}
\restylefloat{table}
\newcommand{\Rnum}[1]{\uppercase\expandafter{\romannumeral #1\relax}}
\usepackage{natbib}
\begin{document}

\title{Unravelling the Complex Structure of AGN-driven Outflows III. the outflow size-luminosity relation}
\author{Daeun Kang$^{1}$ }
\author{Jong-Hak Woo$^{1}$\altaffilmark{$\dagger$}}

\affil{$^{1}$Astronomy Program, Department of Physics and Astronomy, Seoul National University, Seoul 151-742, Republic of Korea} 
\altaffiltext{$\dagger$}{Author to whom any correspondence should be addressed: woo@astro.snu.ac.kr}

\begin{abstract}
Energetic gas outflows driven by active galactic nuclei (AGNs) are considered as one of the mechanisms, by which supermassive black holes affect their host galaxies. To probe the impact of AGN-driven outflows, it is essential to quantify the size of the region under the influence of outflows. In the third of a series of papers, we present the spatially-resolved kinematics of ionized gas for 3 additional type 2 AGNs based on the Gemini Multi-Object Spectrograph (GMOS) integral field spectroscopy. Along with the 6 AGNs presented in our previous works and the 14 AGNs with available GMOS IFU data, we construct a sample of 23 luminous type 2 AGNs at z $<$ 0.2, and kinematically measure the size of ionized gas outflows by tracing the radial decrease of the velocity dispersion of the \oiii\ $\lambda$5007 emission line. The kinematically-measured outflow size ranges from 0.60 to $\sim$ 7.45~kpc, depending on AGN luminosity. We find that the size of the photoionized region is larger than the kinematically-measured outflow size, while the flux-weighted photoionization size is significantly smaller. Thus, using the photoionization size as a proxy for the outflow size will lead to overestimation or underestimation, and introduce a large uncertainty of the mass outflow rate and the energy output rate.
We report the outflow size-luminosity relation with a slope of 0.28$\pm$0.03, which is shallower than the slope of the correlation between the photoionization size and luminosity.  
\end{abstract}
\keywords{galaxies: active - galaxies: kinematics and dynamics - quasars: emission lines}

\section{Introduction}
\label{sec:introduction}    
    
 Active galactic nuclei (AGN) are most powerful energy generators in the universe, inducing large-scale phenomena, i.e., gas outflows and radio jets, which may extend from the central pc region to over kpc-scales~\citep[e.g.,][]{Nesvadba+06, Nesvadba+11, Maiolino+12, Cicone+14, Harrison+14, Liu+14, Husemann+16, Fischer+17}. Whether these galactic-scale phenomena are responsible for connecting the growth of supermassive black holes and their host galaxies has been a contentious issue in the context of galaxy evolution and AGN feedback. The AGN feedback is observationally motivated by the empirical scaling relations between black hole mass and the properties of inactive and active galaxies~\citep[e.g.,][]{Magorrian+98, Ferrarese+00, Gebhardt+00, Marconi&Hunt03, Gultekin+09, Woo+10, McConnell&Ma13, Woo+13, Woo+15}, which may be established by the self regulation between black holes and their host galaxies~(\citealt{Silk&Rees98, Ciotti&Ostriker07, DeGraf+14}, see \citealt{Fabian12}, \citealt{Kormendy&Ho13}, and \citealt{King&Pounds15} for review). 
   
One of the main tasks for ensuring whether gas outflows are suitable as an AGN feedback mechanism is to investigate and quantify how energetic these outflows are and how far outflows can extend to impact on ISM~(see \citealt{Harrison+18} for review). Ionized gas outflows formed in radiatively energetic AGNs are frequently observed by absorption lines, particularly in the X-ray and UV \citep[e.g.,][]{Crenshaw+99, Moe+09, Tombesi+10a, Borguet+12}, and by emission lines in the $\sim$1-10 kpc scale narrow line region (NLR), which is the interface between AGNs and host galaxies~\citep[e.g.,][]{Crenshaw+10, Muller+11, Greene+11, Fischer+13}. Particularly, the strong \oiii\ $\lambda$5007 line has been popularly utilized as a tracer of the ionized gas outflows as the outflow kinematics are well represented in the \oiii\ line profile by asymmetric broad wing components.
   Several studies investigated \oiii\ kinematics for understanding AGN-driven outflows, focusing on individual AGNs \citep[e.g.,][]{Crenshaw&Kraemer00b, Crenshaw+10, Nesvadba+11, Villar-Martin+11, Fischer+13, Karouzos+16a, Karouzos+16b, Bae+17, Revalski+18}, or using a large sample \citep{Boroson05, Greene&Ho05, Zhang+11, Mullaney+13, Bae&Woo14, Wang+18}. These studies demonstrate that outflows are frequently observed in both type 1 and type 2 AGNs. Using a large sample of SDSS type 2 AGNs, \cite{Woo+16} reported that gas outflows manifested by \oiii\ are ubiquitous in luminous AGNs with a trend that the higher the Eddington ratio the stronger gas outflows are \citep[for the discussion on the star formation rate, see][]{Woo+17}. By utilizing the same large sample, \cite{Kang+17} also presented that the outflow kinematics based on the \ha\ line showed a similar trend.
   
While it is clear that outflows are prevalent in AGNs, the size of gas outflows is yet to be properly constrained. For a relatively small number of AGNs, there have been various attempts to measure the size of the photoionized region based on the morphology and distribution of the \oiii\ emission line, using narrow-band images or long-slit spectroscopic data \citep[e.g.,][]{Bennert+02, Schmitt+03, Greene+11}. More recent studies utilized integral field spectroscopy (IFS), which opened a new horizon by making it possible to probe the detailed 2-dimensional structure and kinematics of the NLR~\citep[e.g.,][]{ Liu+13a, Liu+13b, Liu+14, Harrison+14, Husemann+14,  Brusa+16, Karouzos+16a, Karouzos+16b,Bae+17}.
  
However, the size of the outflows has not been well determined because the most previous studies used the flux distribution of the ionized gas rather than the kinematic information. While the flux distribution of ionized gas can provide the size of the photoionization, the photoionization size is not same as the outflow size because outflows may not extend as much as ionizing photons. In other words, the outflow size can be much smaller than the size of the photoionized region, if the kinetic energy does not propagate as efficient as the ionizing photons. Thus, it is more appropriate to measure the size of outflows based on the spatially resolved kinematics. For example, \cite{Karouzos+16b} reported that the outflow velocity and velocity dispersion are radially decreasing, and the size of outflows based on the kinematics is different from that based on the flux distribution of ionized gas \citep[see also][]{Bae+17}.

In this paper, we investigate the size of outflows based on the spatially resolved 2-dimensional kinematics, using a sample of 23 luminous type 2 AGNs, for which the GMOS IFU data are available from our own observations or from the Gemini archive. Nine AGNs were observed over 2 semesters by our Gemini programs, while the other 14 AGNs were observed and presented by \cite{Harrison+14}. In Section~\ref{sec:sample_observation}, we provide the sample selection criteria and in Section~\ref{sec:methodology} we describe how gas kinematics are measured. In Section~\ref{sec:emssion line} we focus on the detailed gas kinematics of the 3 AGNs. Spatial distributions of the outflow kinematics of 23 galaxies are presented in Section~\ref{sec:outflow}. Finally, we discuss our results in Section~\ref{sec:discussion}, and summarize them in Section~\ref{sec:summary and conclusion}. In this paper, we adopted $\Lambda$CDM cosmology with cosmological parameters: H$_{0}$ = 70km s$^{-1}$ Mpc$^{-1}$, $\Omega_{m}$ = 0.30, and $\Omega_{\Lambda}$ = 0.70.

\section{Sample and observation}
\label{sec:sample_observation}

\subsection{Sample selection}
        In the previous studies, \cite{Bae&Woo14} statistically analyzed kinematic properties of the ionized gas outflows based on the \oiii\ emission line using a large sample of $\sim$23000 type 2 AGNs at z $<$ 0.1 selected from the Sloan Digital Sky Survey Data Release 7 (SDSS DR7; \citealt{Abazajian+09}), and \cite{Woo+16} extended the number of samples to $\sim$39000 by including AGNs with a higher redshift out to z$\sim$0.3. Among these type 2 AGNs, we selected a small sample of AGNs with strong outflow signatures in order to spatially investigate outflows and star formation using integral field spectroscopy. To select energetic AGNs with strong outflows, we first applied a luminosity cut as the extinction corrected \oiii\ luminosity L$_{\oiii, cor}$ > 10$^{42}$~\ergs. Then, we selected AGNs based on strong outflow features, i.e., \oiii\ velocity dispersion $\sigma_{\oiii}$ > 350~\kms, or velocity shift |V$_{\oiii}$| > 200~\kms.
We also set the redshift limit as z $<$ 0.1, for securing at least a sub-kpc scale spatial resolution. Using these criteria, we selected 29 AGNs for follow-up studies with IFU. Six of the 29 AGNs were observed in 2015A with the GMOS-N IFU as presented by \citet{Karouzos+16a, Karouzos+16b}. Using the same telescope and instrumental set up, we observed 3 additional targets in 2015B. Note that the method of determining velocity shift has been modified since \cite{Woo+16}, by measuring the velocity of the line based on the flux weighted center (i.e., the first moment, see Eq.~\ref{eq:fmom}), instead of the peak of the line \citep{Bae&Woo14}. For this reason, the velocity shifts of \oiii\ measured by the new method were slightly changed  \citep[see][]{Woo+16}. 
   
\begin{figure}
	\center
	\includegraphics[width=0.46\textwidth]{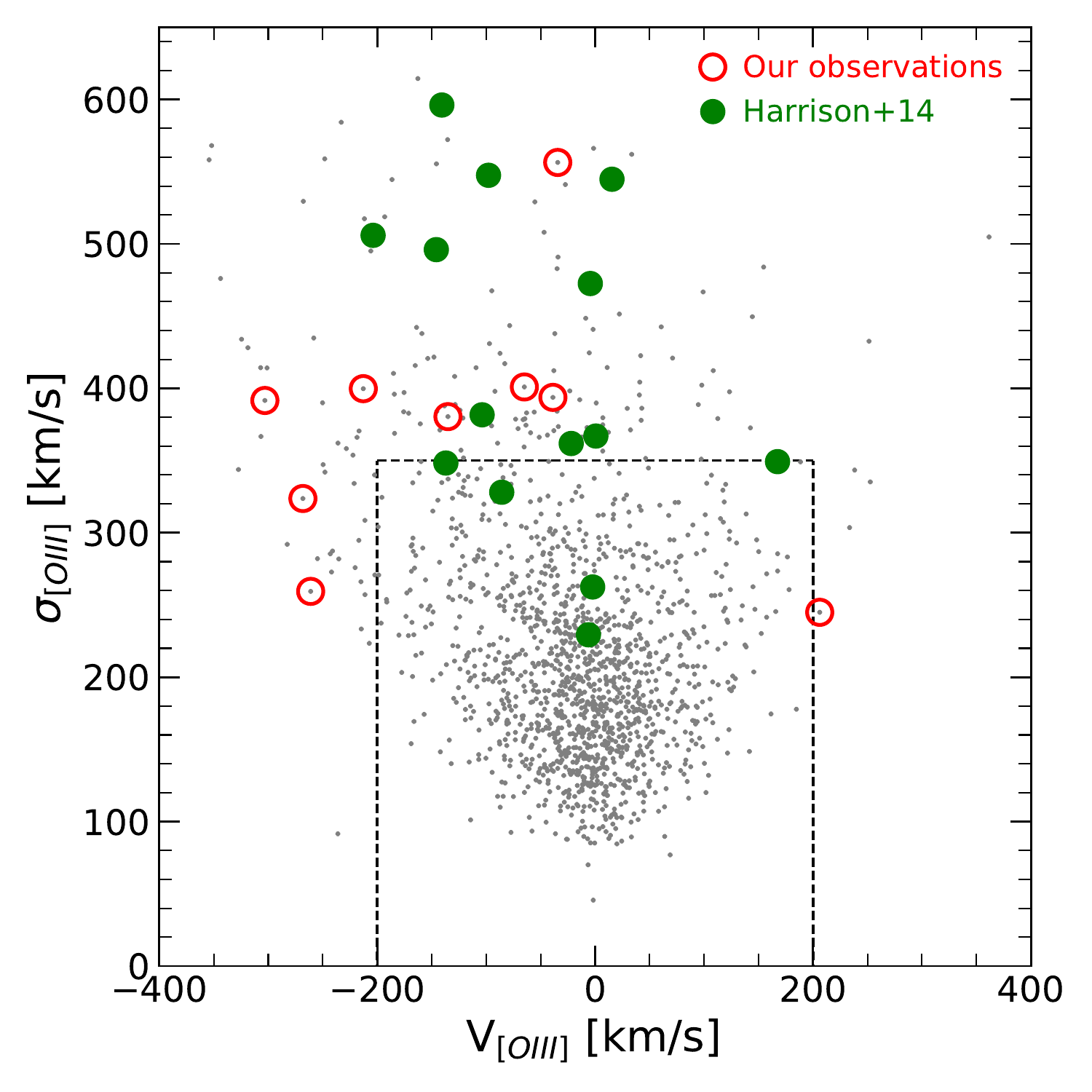}
	\caption{Velocity-velocity dispersion diagram (VVD) of luminous SDSS AGNs (i.e., L$_{\oiii, cor}$ > 10$^{42}$~\ergs) from \cite{Woo+16} (black dots)  along with the 9 AGNs in our GMOS programs (red open circles) and the 14 AGNs observed by \cite{Harrison+14} (green filled circles). Dashed lines indicate the selection criteria for AGNs with strong outflows (i.e., $\sigma_{\oiii}$ > 350~\kms, or |V$_{\oiii}$| > 200~\kms).   \label{fig:vvd} } 
\end{figure}

        In addition to the nine type 2 AGNs observed with our GMOS programs, we selected additional 14 type 2 AGNs from the Gemini archive, which were observed using the GMOS-South IFU in 1 slit mode with the B1200 grating by \cite{Harrison+14}. Note that this setup is similar to ours although the spectral range covers only the \hb\ and \oiii\ region.  
    \cite{Harrison+14} selected and observed 16 luminous AGNs (L$_{\oiii}$ > 5 $\times$ 10$^{41}$~\ergs) with a broad component in \oiii\ line profile (i.e., FWHM > 700~\kms) at z < 0.2, of which the \oiii\ properties are quite similar to those of our sample (see \citealt{Harrison+14} for the detailed selection criteria). We only utilize 14 of them by excluding 2 AGNs, namely, J1316+1753, J1356+1026, which have clear and luminous double peaks in their \oiii\ line profile. For the case of J1356+1026, it is shown that there is an ongoing merger of two distinct type 2 AGNs \citep{Greene+12, Harrison+14}. For the case of J1316+1753, although it was not confirmed as a dual AGN by \cite{Harrison+14}, the \oiii\ line profile has a complex nature, requiring multiple Gaussian components with two separate velocity centers. Since the ionized gas kinematics of this AGNs is clearly different from that of the other AGNs, we exclude this target from our analysis. 
Kinematic properties of the combined sample are presented in Figure~\ref{fig:vvd}. 
    
\renewcommand{\arraystretch}{1.25}
\begin{deluxetable}{ccccc}[t]
   	\tablewidth{0.45\textwidth}
    \tablecolumns{5} 
    \tablecaption{Observing log of three additional targets \label{table:obs}}
    
    \tablehead
    {
      	\colhead{ID}   &   \colhead{RA}        &   \colhead{Dec}        &   \colhead{Exposure}  &   \colhead{Seeing}    \\        
        \colhead{}        &   \colhead{(hh mm ss)}  &   \colhead{(dd mm ss)}   &    \colhead{sec}        &          \colhead{\arcsec} \\ 
        \colhead{(1)}        &   \colhead{(2)}  &   \colhead{(3)}   &    \colhead{(4)}        &          \colhead{(5)}
    }
    \startdata
        J205537-003812 & 20 55 37 & -00 38 12 & 2700 & 0.3  \\ 
        J213333-071249 & 21 33 33 & -07 12 49 & 3420 & 0.5  \\ 
        J214600+111326 & 21 46 00 & +11 13 26 & 3420 & 0.4
    \enddata
    \tablecomments{(1) SDSS ID of targets; (2) R.A (J2000); (3) Decl. (J2000); (4) exposure time; (5) seeing size.}
\end{deluxetable}

\subsection{Observation and data reduction}
 \citet{Karouzos+16a} presented the observation and analysis of the initial sample of six AGNs. Here, we briefly summarize the data reduction process for 3 additional targets. Note that we also reduced the GMOS-S data for the 14 AGNs selected from the Gemini archive. Three additional AGNs  were observed in 2015B using the GMOS-N IFU in 1-slit mode (ID:GN-2015B-Q-92, PI:Woo). To cover a wide wavelength range, including both \oiii\ and \ha\ lines,  we used the B600 grating with a 2-pixel spectral binning, which provided an instrumental resolution $\sigma_{inst} \sim$90~\kms. The field of view (FOV) is 3".5 $\times$ 5".0, corresponding to 3.2 kpc $\times$ 4.6 kpc for the nearest AGN or 11.5 kpc $\times$ 16.4 kpc for the farmost AGN. The size of one spaxel corresponds to 0".1 while the typical seeing was 0".3-0".5 (see Table~\ref{table:obs}).
    
 Preprocessing was performed mainly with the Gemini IRAF package in a standard order, including bias subtraction, flat fielding, and wavelength calibration after the cosmic ray removal by using the PyCosmic routine~\citep{Husemann+12}. Flux calibration was conducted in two steps: first, we used the sensitivity function obtained using a standard star, which was observed in the same observing run. Second, we compared the central spectrum integrated within a 3" diameter aperture with the SDSS spectrum for a consistency check. The strength of the emission-line flux density is consistent between the GMOS and SDSS spectra.

\section{Methodology}
\label{sec:methodology}
\subsection{Emission line fitting}
\label{subsec:method1}

We adopt the same process of measuring kinematic properties of ionized gas as we used for several previous studies~\citep{Woo+16, Karouzos+16a, Karouzos+16b, Bae+17, Eun+17, Kang+17}. First, we fit stellar absorption lines in the 5100-5400\AA\ range (including e.g, the \mgb\ triplet), remove stellar continuum, and measure the systemic velocity of each galaxy, using the pPXF code~\citep{Cappellari&Emsellem04, Cappellari17}.  For the 9 AGN in our Gemini program, we fit stellar continuum in each spaxel, then we determine the systemic velocity by calculating the median of the stellar velocity, using the central 9 spaxels (corresponding to 0".9$\times$0".9). 
In the case of the 14 AGNs from the Gemini archive, which has relatively weak stellar absorption lines, we integrate the spectra within the 3" diameter aperture to fit stellar continuum. Nevertheless, because 10 AGNs do not show detectable absorption lines even in the combined spectra, we determine the systemic velocity using the peak of the \oiii$\lambda$5007 emission line. Note that although the flux-weighted center of the \oiii\ line shows a velocity offset with respect to stellar lines, the peak of the line is expected to be close to the systemic velocity of the host galaxies. Though the peak of the \oiii\ often shows a few tens of \kms\ scale velocity shift \citep[see][]{Bae&Woo14, Karouzos+16a}, this does not affect our results because we mainly use the velocity dispersion for determining the outflow size.
The measured systemic redshifts are presented in Table~\ref{table:prop}.
    
        Second, we fit each emission line using the MPFIT package~\citep{Markwardt09}.  While we fit \ha, \nii\ doublet, \oiii\ doublet, and \hb\ for the 9 AGNs based on our GMOS data, we fit only \hb\ and the \oiii\ doublet for the other 14 AGNs due to the limited spectral range. Each emission line is fitted with a double-Gaussian model, and re-fitted with a single-Gaussian model if the amplitude-to-noise ratio (A/N) of either Gaussian model is smaller than 3, for avoiding unreliable fits. We simultaneously fit each doublet with the fixed flux ratio of 2.993 \citep{Dimitrijevic+07}.    
    
     Third, using the best-fit model, we calculate the 1st and 2nd moments of the line profile as
     \begin{eqnarray} 
     \lambda_{0} = {\int \lambda f_\lambda d\lambda \over \int f_\lambda d\lambda}
     \label{eq:fmom}
     \end{eqnarray}
     
     \begin{eqnarray}
     \Delta\lambda^2 = {\int \lambda^2 f_\lambda d\lambda \over \int f_\lambda d\lambda} - \lambda_0^2.
     \label{eq:smom}
     \end{eqnarray}
     Then, we determine the velocity shift with respect to the systemic velocity, and the velocity dispersion for each emission line. For velocity dispersion, we correct for the instrumental broadening using the instrumental resolution (i.e., $\sigma_{inst}$ $\sim$ 90~\kms\ for the 9 AGN in our observations and 34~\kms\ for the 14 AGNs from the archive).
     Last, we conduct Monte Carlo simulations to quantify the uncertainties of velocity shift and velocity dispersion. We generate 100 mock spectra by randomizing flux using flux error and fit each spectrum to measure velocity shift and velocity dispersion. The standard deviation of the distribution of the measurements is adopted as the 1$\sigma$ error of each property.

\subsection{Stellar velocity dispersion}
\label{subsec:method2}
Stellar velocity dispersion is required in order to separate the outflow component from the gravitational component in a given emission line profile.
For the 9 AGNs in our GMOS observations, we adopt the flux-weighted stellar velocity dispersion within the 3" diameter aperture, using the stellar velocity dispersion measured at each spaxel as
	\begin{eqnarray} 
	\sigma_{*}^{2} = {\int \sigma_{*}^{2}(x,y) F(x,y) dA \over \int F(x,y) dA}.
	\label{eq:svd}
	\end{eqnarray}
	 
 For the majority of the other 14 AGNs, it is difficult to obtain a reliable fit on the stellar component due to the lack of detectable stellar absorption lines even if we use the integrated spectra inside the 3" diameter aperture. Only for 4 AGNs (namely, J1010+0612, J1100+0846, J1216+1417, and J1339+1425), we are able to fit stellar component to obtain systemic velocity and stellar velocity dispersion from the integrated spectra within the 3" diameter aperture. Note that the stellar velocity dispersion measured from the integrated spectra may suffer from  rotational broadening. As a consistency check, we compare the stellar velocity dispersions measured from spatially-resolved spectra (i.e., based on Eq. 3) and those from the integrated spectra using the aforementioned 9 AGNs. We find that the ratio of stellar velocity dispersions is $0.99\pm0.04$ dex, indicating that the effect of the rotational broadening is only a few percent, which varies presumably depending on the inclination and bulge-to-disk flux ratio of individual host galaxies.
  
For the other 10 AGNs, systemic velocity is determined using the peak of the \oiii\ emission line as mentioned in Section~\ref{subsec:method1}, while stellar velocity dispersion is substituted by the velocity dispersion of the narrow component of either \oiii\ or \hb. For 6 AGNs, we use the narrow component of \oiii\ since \hb\ is weak, while for the other 4 AGNs, we are able to use \hb. Several studies showed that the narrow component of emission lines usually traces the gravitational potential~\citep[e.g.,][]{Greene&Ho05, Woo+16, Karouzos+16a, Kang+17}.
Nevertheless, the narrow component of these lines may be influenced by other effect, i.e., gas pressure or outflows, especially in the case of \oiii\ \citep{Karouzos+16a, Karouzos+16b}.
We directly test the difference of gas and stellar kinematics using the 13 AGNs with measured stellar velocity dispersion. 
The velocity dispersion of the narrow component of \oiii\ is larger than stellar velocity dispersion by a factor of $1.09\pm0.32$, while the velocity dispersion ratio between \hb\ and stellar lines is $0.92\pm0.30$. If we compare them in log scale, the mean \oiii-to-stellar velocity dispersions is $1.01\pm0.06$ and the mean \hb-to-stellar velocity dispersion is $0.98\pm0.06$. We further test the difference of gas and stellar kinematics, using a sample of $\sim$80 SDSS type 2 AGNs with strong outflow signatures, which are selected with the same criteria used for the GMOS sample selection. The mean \oiii-to-stellar velocity dispersion ratio of this sample is $0.99\pm0.11$ in log scale, while the mean \hb-to-stellar velocity dispersion ratio is $0.93\pm0.07$, indicating that the difference is insignificant compared to the measurement uncertainties. 
If we use the mean ratio from the 13 AGNs as a correction factor (i.e., 1.09 and 0.92, respectively for \oiii\ and \hb), the outflow size and size-luminosity relation in the following analysis slightly change, however, the effect is negligible. Thus, we decide not to apply the correction factor, and use the velocity dispersion of \oiii\ or \hb\ as a proxy for stellar velocity dispersion for the 10 AGNs, for which stellar velocity dispersion is not directly measured.

\section{Emission-line properties in the NLR : \oiii\ $\&$ \Ha}
\label{sec:emssion line}
	In this section, we investigate the spatial distributions of the emission-line properties. We only present the results for the 3 additional AGNs observed in 2015B, as the results of the other 20 AGNs were already presented in the previous studies \citep{Karouzos+16a, Karouzos+16b, Harrison+14}. We mainly present the results based on \oiii\ and \ha\ in Figure 2 to Figure 5. 

 \begin{figure}
	\center
	\includegraphics[width=0.49\textwidth]{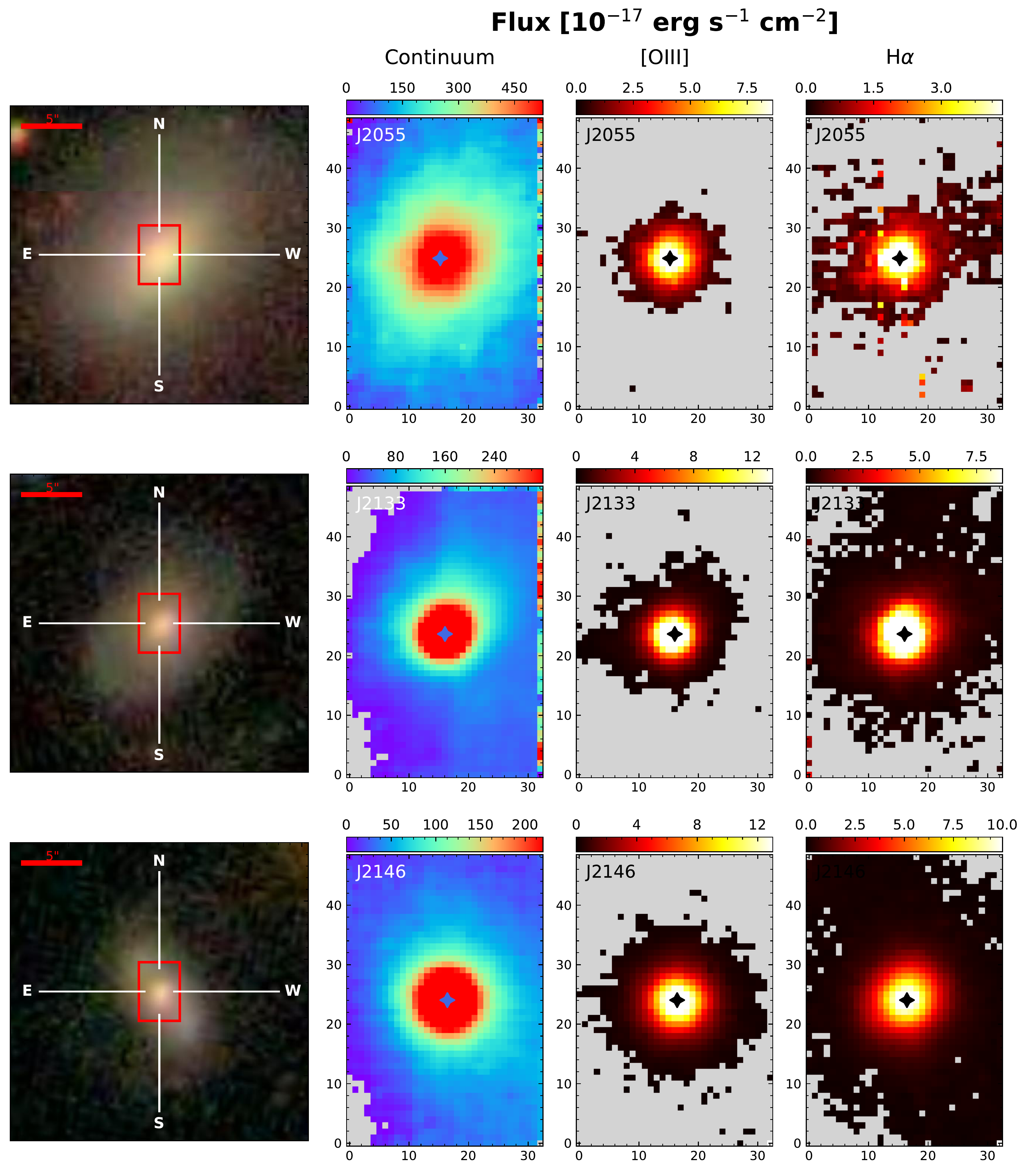}
	\caption{First column: SDSS gri composite images. The FOV of GMOS IFU (3".5x5".0) is denoted with the red rectangle,
	 while the horizontal red bar at the upper left side indicates a 5".0 scale. Second column: Continuum flux maps integrated over the full wavelength range. Third $\&$ fourth columns: flux maps of \oiii\ and \ha, respectively. One pixel corresponds to 0".1.
	Gray color indicates the spaxels where continuum or emission line is weak or non-detected.
	The center of each galaxy based on the continuum flux distribution is designated by a cross.
		\label{fig:ftotal} } 
\end{figure}

\begin{figure}
	\center
	\includegraphics[width=0.48\textwidth]{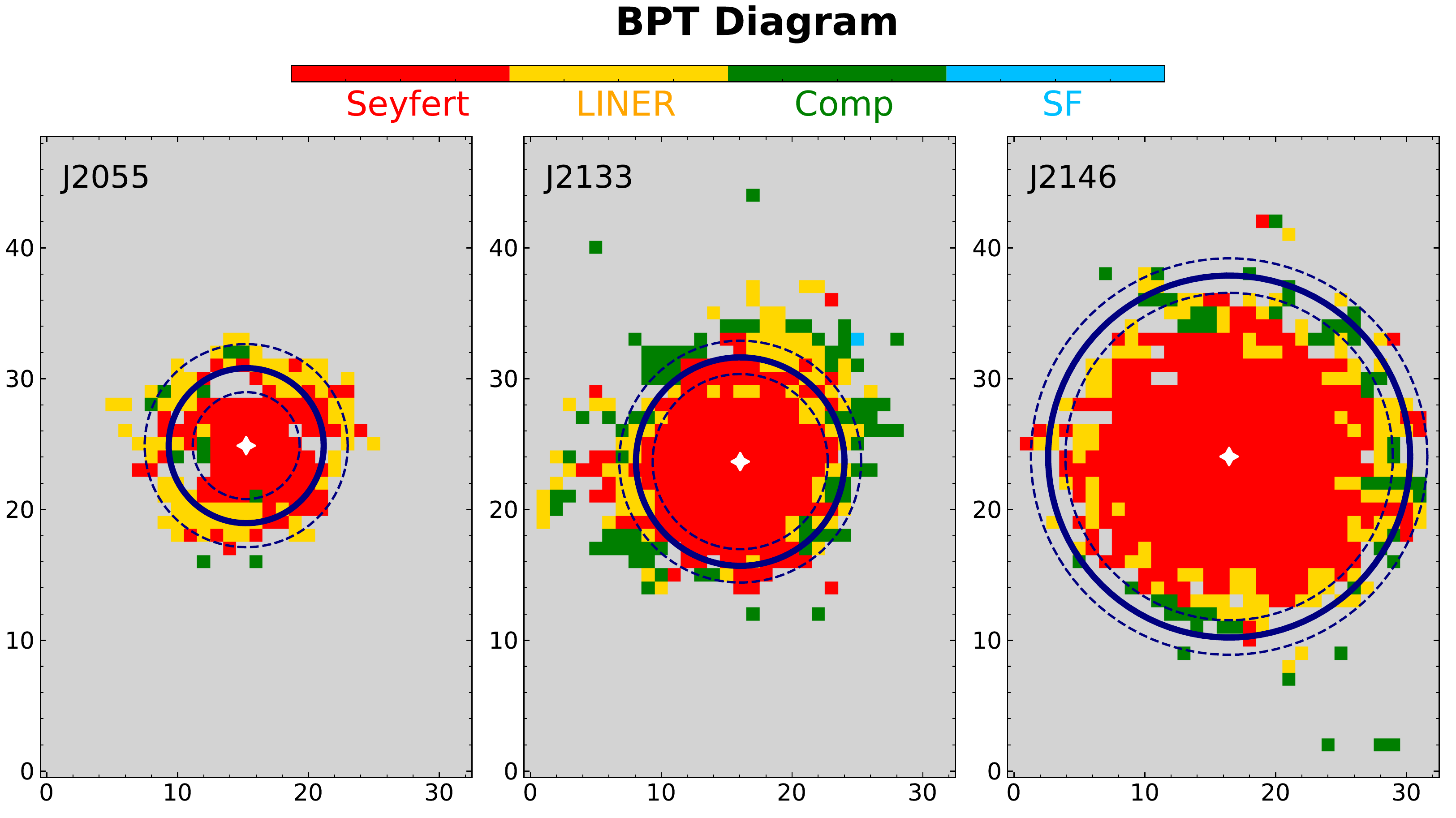}
	\caption{BPT diagram maps. Color code represents each classification: Seyfert (red), LINER (yellow), composite (green), star-forming region (blue). Classification was not performed if the four emission lines are weak or non-detected with \ha\ S/N < 3, or \oiii\ S/N < 3, or \nii\ $\lambda$6583 < 3, or \hb\ S/N < 1 (grey spaxels). The size of outflows and the 1 $\sigma$ uncertainty are with black solid and dashed lines (see Table~\ref{table:prop}) while the center of the continuum flux distribution is denoted with a cross.
			\label{fig:bpt}} 
\end{figure}

\begin{figure*}
	\center
	\includegraphics[width=0.99\textwidth]{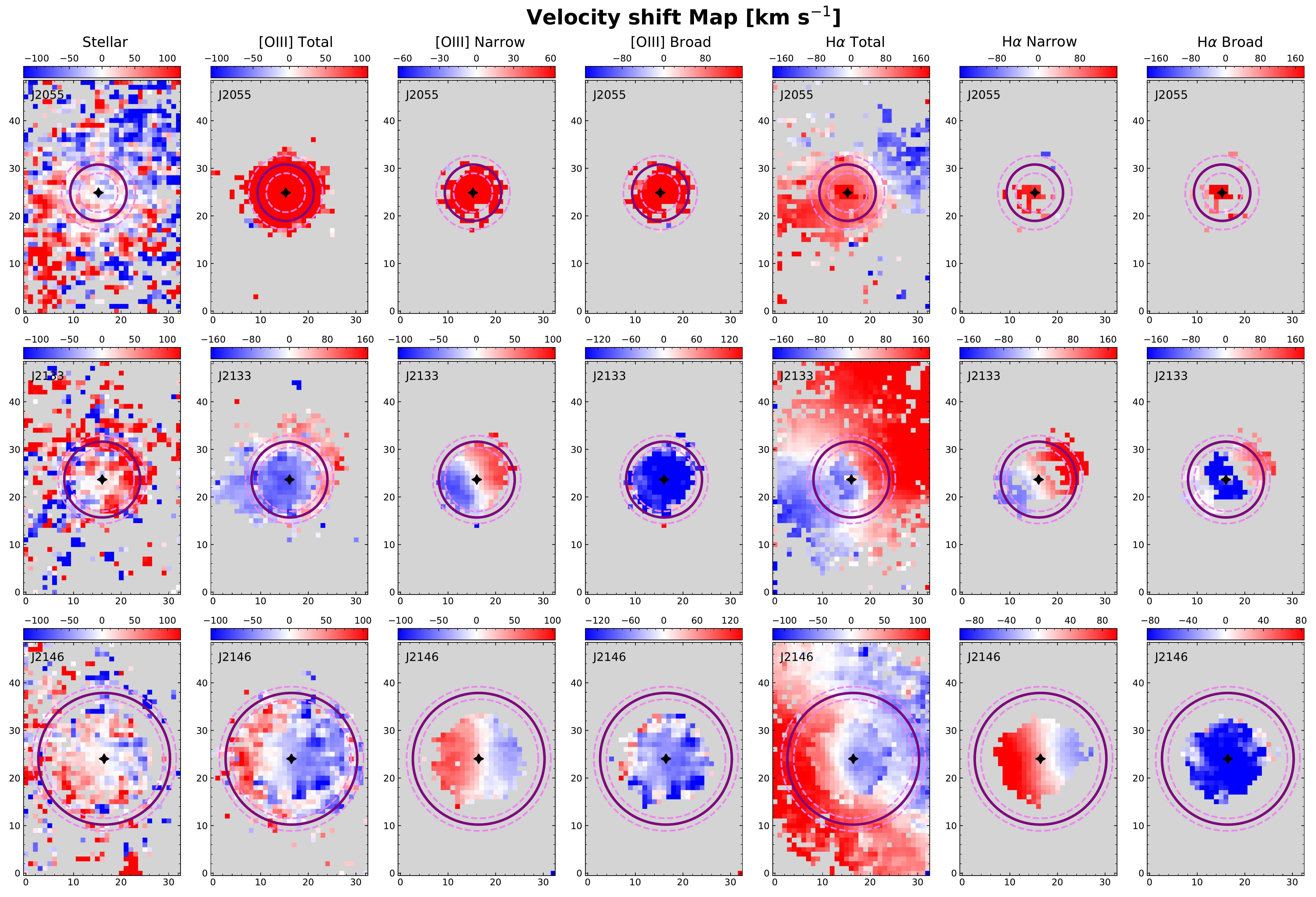}
	\caption{Velocity maps of each object measured from stars (1st column), \oiii\ based on the total (2nd column), narrow (3rd column), broad (4th column) components of the line profile, and \ha\  based on the total (5th column), narrow (6th column), broad (7th column) components of the line profiles. Gray color indicates the spaxels where continuum (1st column) or emission line (the other columns) is weak or non-detected (i.e, S/N < 3).  Circles and symbols are the same as in Figure~\ref{fig:bpt}.
\label{fig:vtotal}} 
\end{figure*}

\begin{figure*}
	\center
	\includegraphics[width=0.98\textwidth]{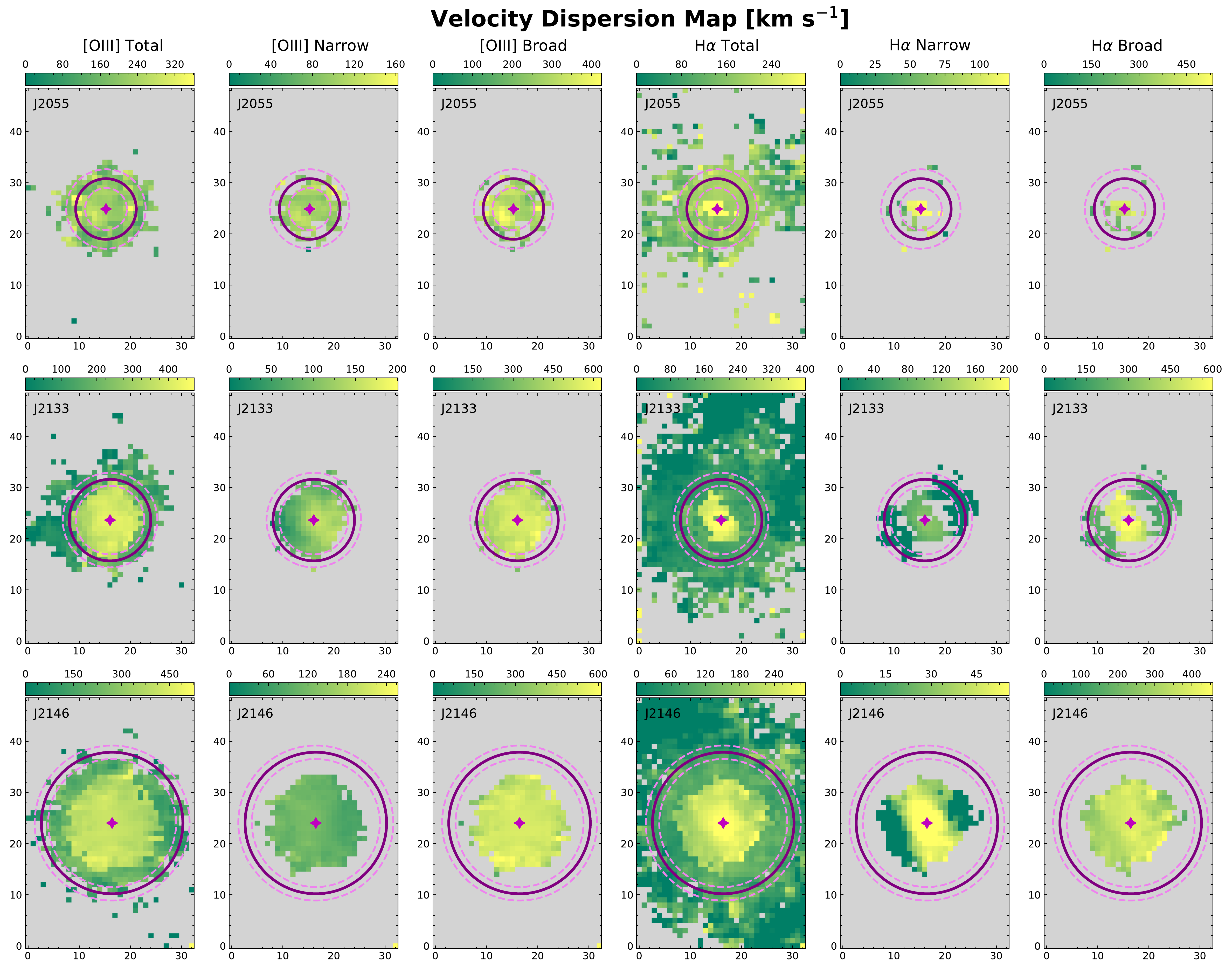}
	\caption{Velocity dispersion maps of \oiii\ based on the total (1st column), narrow (2nd column), and broad (3th column) components of the line profile, and \ha\  based on the total (4th column), narrow (5th column), broad (6th column) components of the line profiles. Gray color indicates the spaxels where emission line is weak or non-detected (i.e, S/N < 3).  Circles and symbols are the same as in Figure~\ref{fig:bpt}.	
\label{fig:stotal}} 
\end{figure*}

\subsection{Flux and Flux ratio}
\label{subsec:emssion line flux}

We present the SDSS gri composite images and spatial distributions of continuum and emission line fluxes in Figure~\ref{fig:ftotal}. Note that gray spaxels in 2-dimensional maps indicate the failure of emission-line fitting due to several reasons, including low signal-to-noise ratio (i.e., S/N < 3), or velocity dispersion smaller than the spectral resolution limit. The continuum flux maps show a good agreement with the SDSS images, and the \oiii\ and the \ha\ flux maps are similar to the continuum flux map with a consistent flux center of the continuum and emission line fluxes. 
  
	To investigate the photoionization properties, we calculate the flux ratios using the 4 emission-lines, i.e., \ha, \oiii$\lambda$5007, \nii$\lambda$6584, and \hb, and classify each spaxel into 4 categories, namely, Seyfert, Low-Ionization Nuclear Emission-line Region (LINER), composite and star-forming galaxies in the emission-line ratio diagram~\citep{Baldwin+81, Kauff+03, Kewley+06}
	as shown in Figure~\ref{fig:bpt}. Note that we only use spaxel with an enough S/N ratio (i.e., S/N > 3 for \oiii$\lambda$5007, \ha, \nii$\lambda$6584, and S/N >1 for \hb), while we separate the Seyfert region from the LINER region if \oiii/\hb\ > 3, which is to be consistent with our previous analysis \citep{Karouzos+16b, Bae+17, Kang+17}.

	For all 3 objects, the center is dominated by AGN photoionization, while LINER and composite regions are located at the boundary of Seyfert region (see Figure~\ref{fig:bpt}). The edge of the Seyfert region coincides with the edge of the outflow region (navy large circle in each map) for J2055 and J2133 (see Section 5 for the outflow size), while in the case of J2146, the size of the outflow region is slightly extended to the edge of the LINER region. These trends are consistent with those of other type 2 AGNs presented by~\cite{Karouzos+16a, Karouzos+16b}.

\subsection{Kinematics of ionized gas}

\begin{figure*}
	\includegraphics[width=0.33\textwidth]{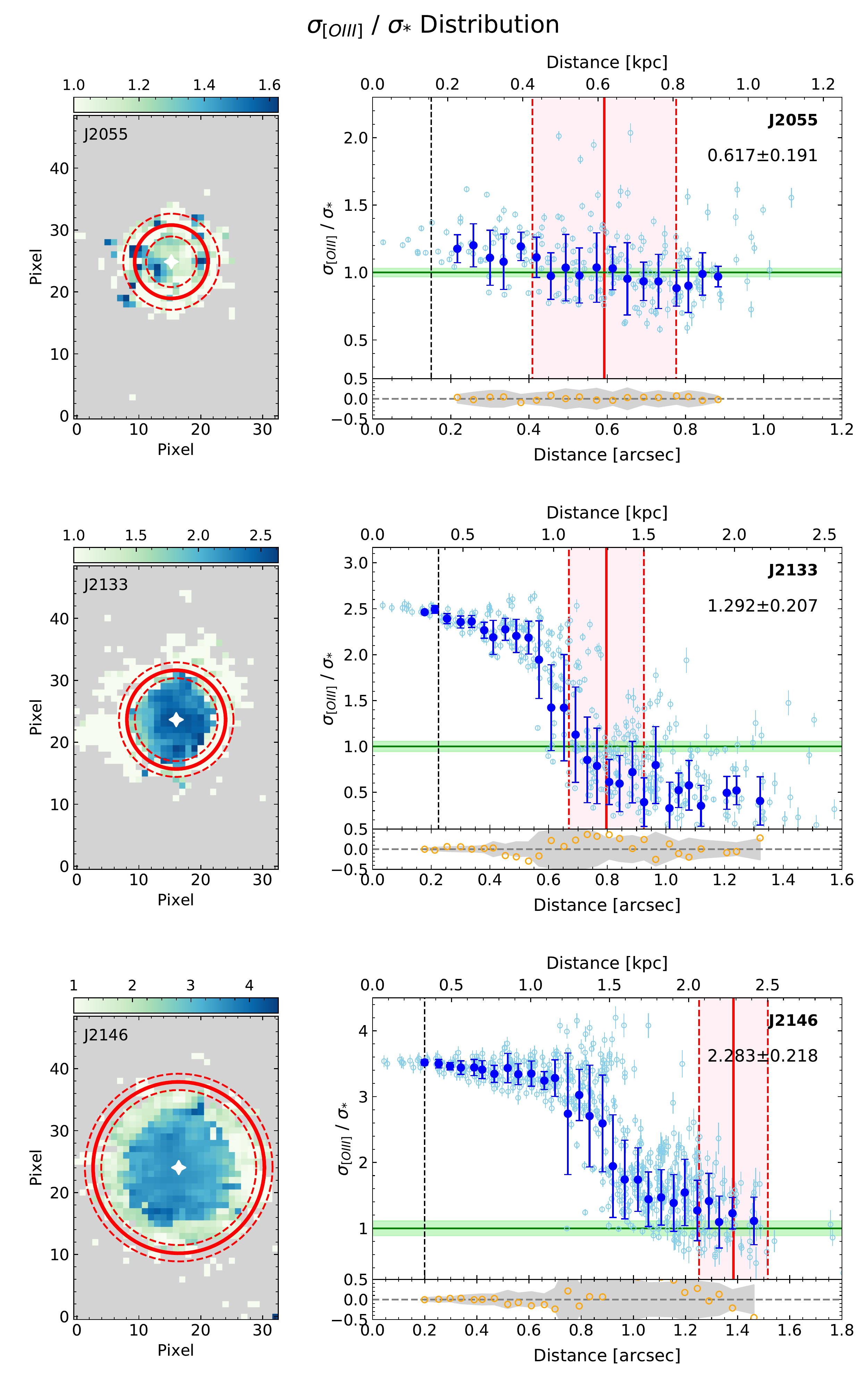}
	\includegraphics[width=0.66\textwidth]{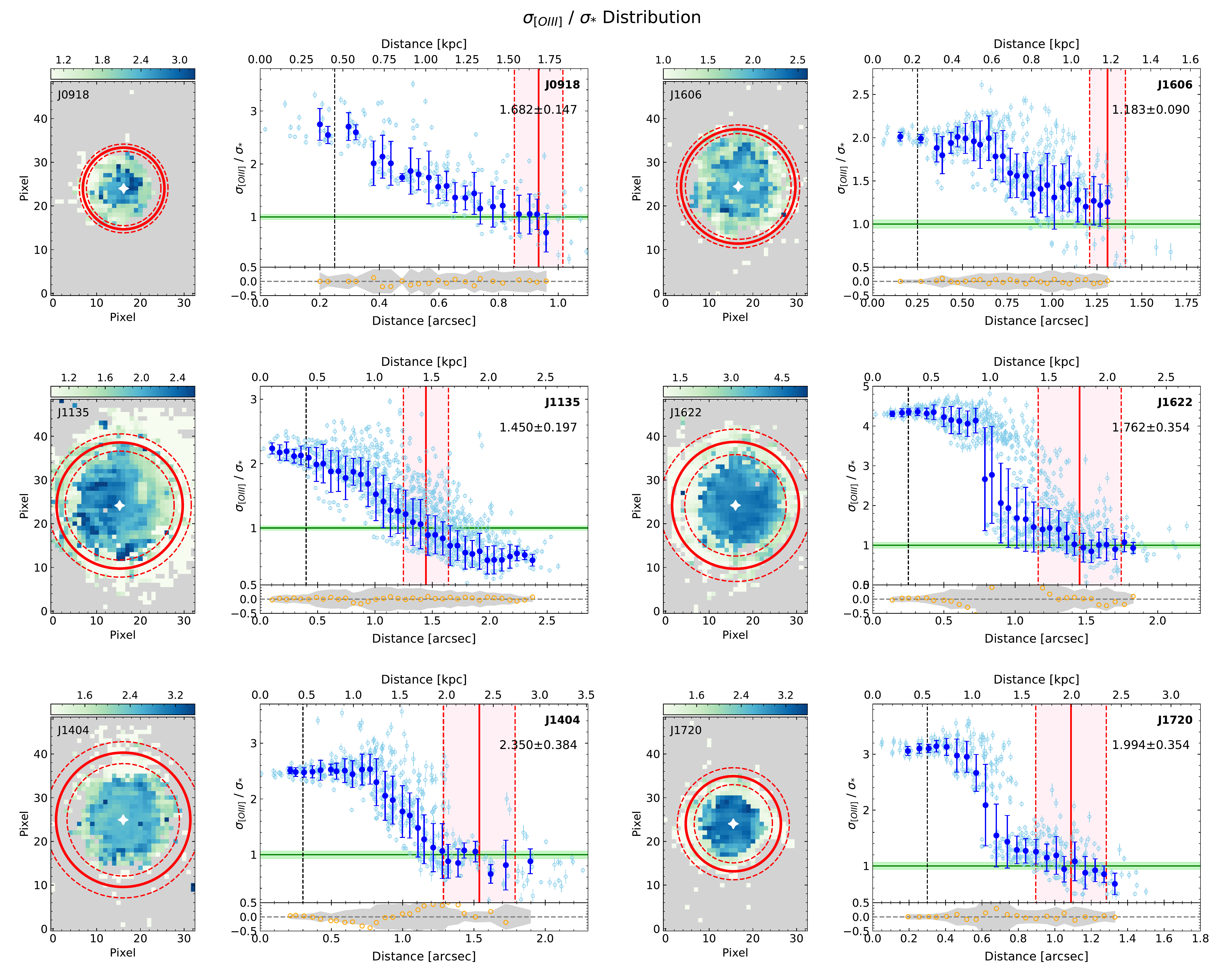}
	\caption{Two-dimensional spatial distribution (left) and radial profile of \oiii\ velocity dispersion normalized by stellar velocity dispersion (right) for each of 9 AGNs in our GMOS programs. In the 2-d maps, the kinematically measured outflow size is indicated by the red circle with 1$\sigma$ error, while gray color indicates unusable spaxels. The color denotes the ratio of \OIII\ velocity dispersion to stellar velocity dispersion. 
		In the radial profile, the error-weighted mean (large blue circles) of the measurements of each spaxel (light-blue points) is  presented as a function of the radial distance. The outflow size is defined when the \OIII-to-stellar velocity dispersion ratio becomes unity (red solid line). The range of the outflow boundary is represented by the vertical pink area surrounded by two vertical red dashed lines. The 1$\sigma$ error range of the velocity dispersion ratio (green box) is used to determine the 1$\sigma$ error of the outflow size. The seeing size (half of FWHM) is denoted with black dashed lines. Numbers at the upper left side of each plots indicate the measured outflow radii and their uncertainties before subtracting the seeing size.
		Orange circles below the radial profile show residual of the best-fit polynomial of the radial profile and the gray area which encloses the orange circles represents the 1$\sigma$ uncertainty of the averaged velocity dispersion ratio at each distance bin (error bars of filled blue circles). 
 \label{fig:outflow1}} 
\end{figure*}

\begin{figure*}
	\includegraphics[width=1.0\textwidth]{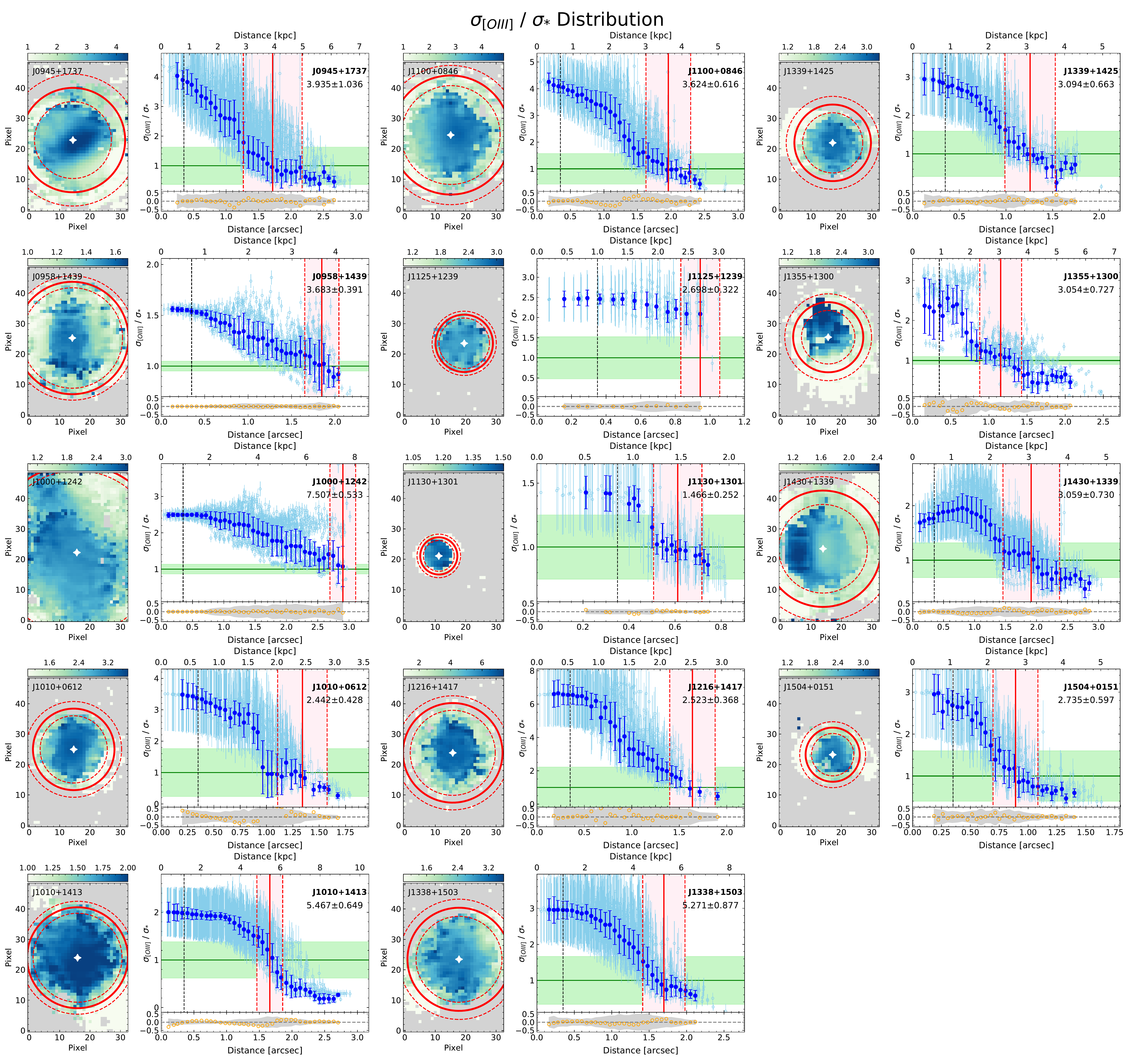}
	\caption{Two-dimensional spatial distributions and radial profiles of the velocity dispersion ratio for the 14 AGNs from the archive. Symbols are same as in Figure~\ref{fig:outflow1}. 	\label{fig:outflow2}} 
\end{figure*}

\begin{figure*}
   	\includegraphics[width=0.33\textwidth]{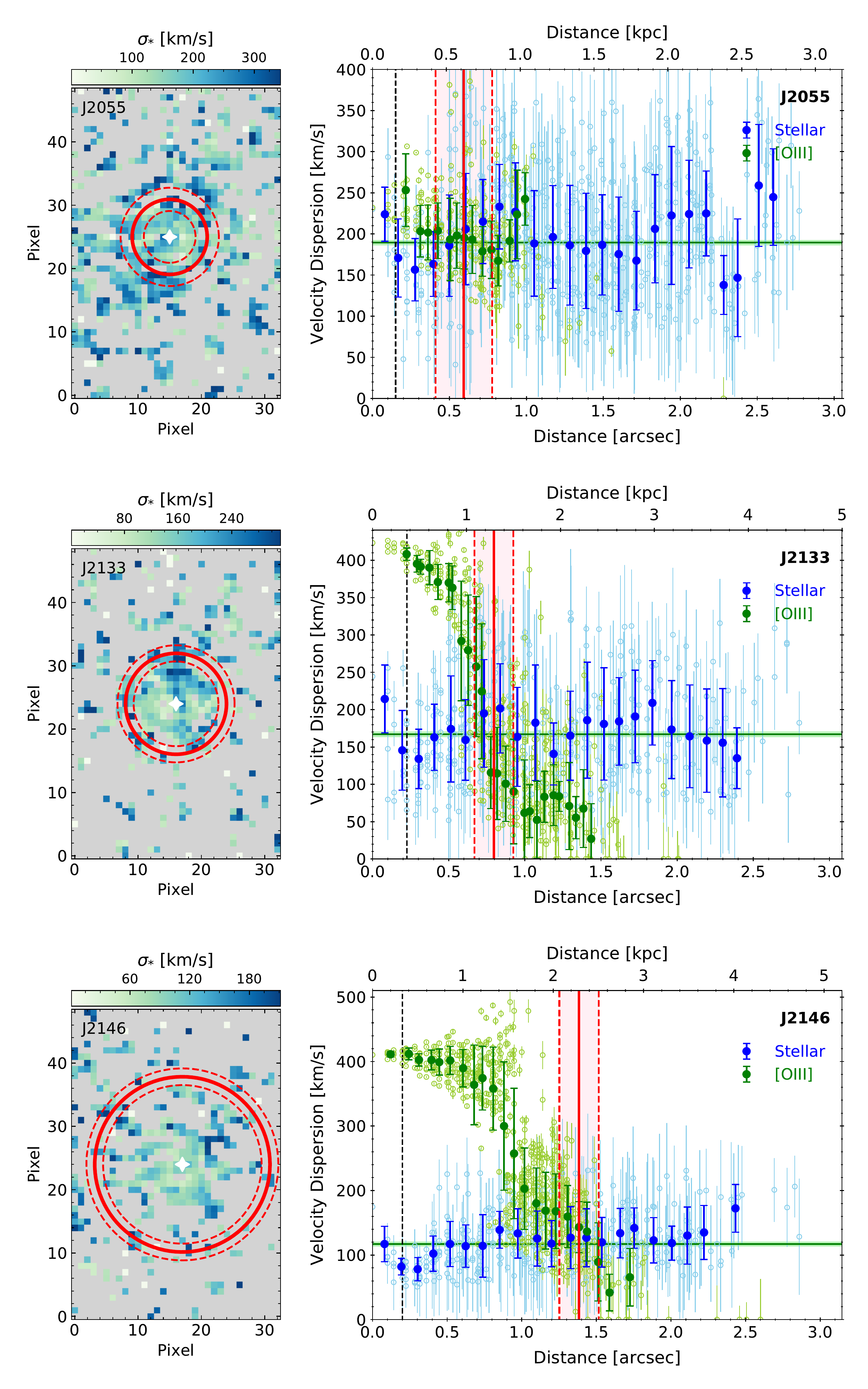}
    \includegraphics[width=0.66\textwidth]{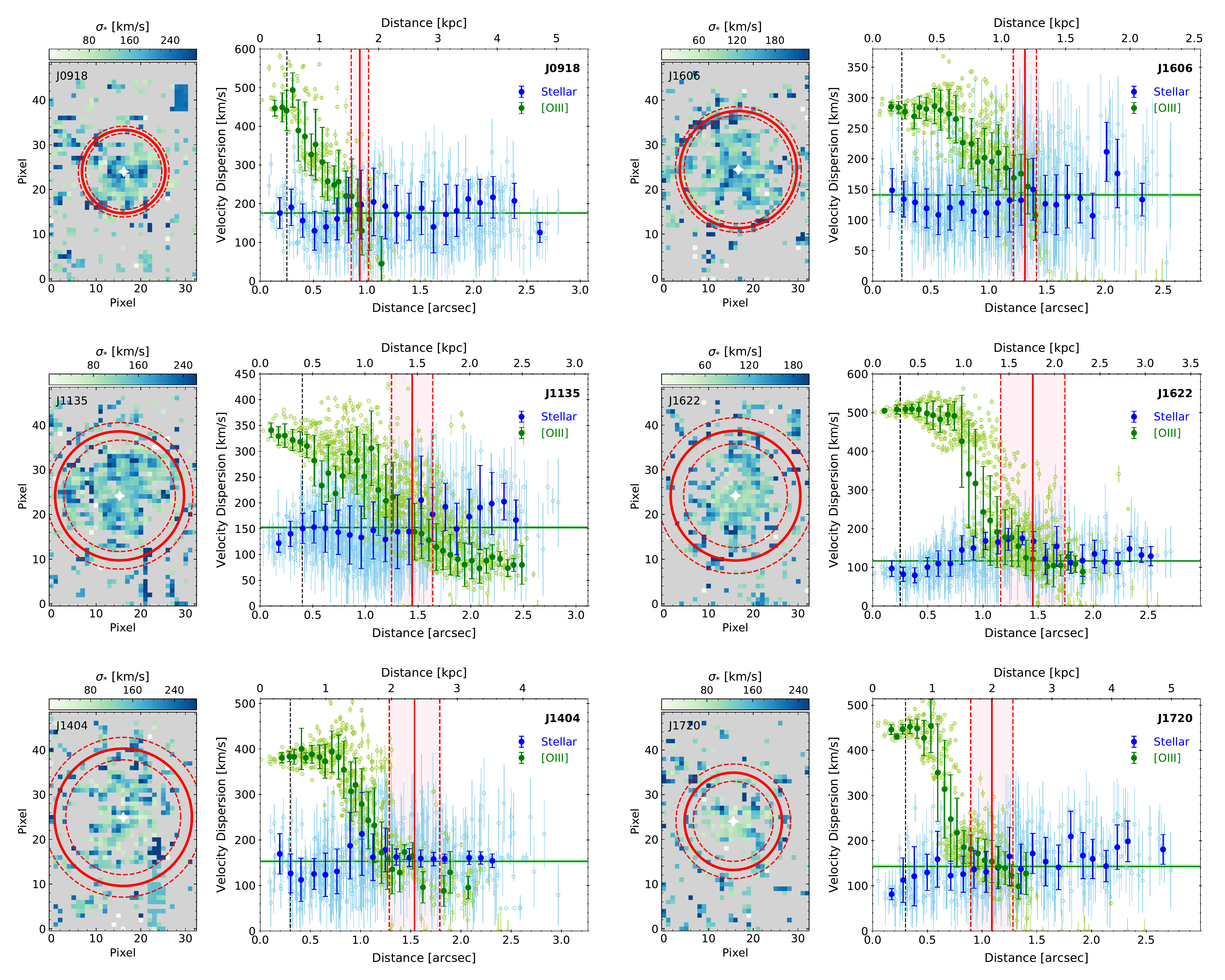}
	\caption{Two-dimensional spatial distribution of stellar velocity dispersion (left) and the radial profile of the velocity dispersions (right) based on stars (blue) and \oiii\ (green) for each of 9 AGNs in our GMOS programs. Symbols are same as in Figure 6. The green horizontal line and box indicates the flux-weighted stellar velocity dispersion and 1 $\sigma$ error
	based on the measurement at each spaxel within 3\arcsec\ diameter aperture. 
        \label{fig:svd}} 
\end{figure*}

We investigate the spatial distribution of gas kinematics, using \oiii\ and \ha.
First, we present the velocity maps of stars and gas in Figure 4. Stellar velocity maps show a rotation in all 3 galaxies, although the measurements are relatively uncertain due to the low S/N of the stellar component. 
In the case of ionized gas, we measure gas velocity using the total line profile, the broad component, and the narrow component, respectively. Note that velocity maps of both broad and narrow components only show the spaxels of which emission line profiles are well decomposed into two Gaussian components. The velocity map based on the total line profile represents the combination of two different gas kinematics: one reflects the gravitational potential of the host galaxy and the other manifests AGN-driven outflows. In J2133 and J2146, for example, gas velocity at the center indicates outflows (blueshift), while the outer part shows a rotation.

A broad component of each emission line is detectable only at the central part, and both broad \oiii\ and \ha\ components show outflows in the same direction, indicating the same non-gravitational influence is manifested. J2055 exhibits receding outflows (i.e., redshifted) at the center, while J2133 and J2146 reveal approaching outflows (i.e., blueshifted). The projected maximum velocity shift of the broad component ranges between 200$\sim$250~\kms.
	The velocity map of the narrow components of both emission lines in J2133 and J2146 reveals a rotation in the same direction as in the stellar velocity map, suggesting that the narrow component, especially in \ha, generally follows the gravitational potential of the host galaxies~\citep{Woo+16, Karouzos+16a, Kang+17}.
Note that while the velocity of the narrow component of \oiii\ exhibit a rotation in J2133 and J2146, the narrow \oiii\ components do not show a galactic rotation in many other AGNs~\citep{Karouzos+16a, Bae+17}. \cite{Woo+16} also showed that for some AGNs, a discernible velocity shift is detected in the narrow component as well as in the broad component. 

Second, we investigate the spatial distribution of gas velocity dispersion in Figure~\ref{fig:stotal}.
Both \oiii\ and \ha\ velocity dispersions measured from the total line profile have their maximum value at the central part. For example, the central \oiii\ velocity dispersion is $\sim$450~\kms\ in J2133 and J2146, and 350~\kms\ in J2055, while the distribution of \ha\ velocity dispersion shows smaller amplitude than that of \oiii, with the maximum velocity dispersion ranging between 300$\sim$400~\kms.
	Velocity dispersion of the narrow component in both emission-lines is similar to the stellar velocity dispersion, which ranges between 140 and 190~\kms, suggesting that the narrow component follows the gravitational potential of the host galaxy. 
	In contrast, the broad component, which represents the non-gravitational potential, shows considerably large velocity dispersion compared to the velocity dispersions measured from the narrow component or the total line profile. For example, the velocity dispersion of the broad component in \oiii\ ranges up to $\sim$600~\kms, while  the velocity dispersion of the broad component in \ha\  shows a similar or slightly smaller range.

\renewcommand{\arraystretch}{1.25}
\begin{deluxetable*}{ccccccc}[t]
	\tablewidth{0.98\textwidth}
	\tablecolumns{7} 
	\tablecaption{Galaxy and emission-line properties \label{table:prop}}
	
	\tablehead
	{
		\colhead{Name}  &   \colhead{z}  &   \colhead{$\sigma_{*}$}     &   \colhead{log (L$_{\oiii}$)}    &  \colhead{V$_{\oiii}$}  & 
		\colhead{$\sigma_{\oiii}$}  &   \colhead{Outflow size}    \\        
		\colhead{}       &   \colhead{}    &   \colhead{(\kms)}  &   \colhead{(\ergs)}   &    \colhead{(\kms)} &    \colhead{(\kms)}        &          \colhead{(kpc)} \\
		\colhead{(1)}           & \colhead{(2)}       &   \colhead{(3)}    &   \colhead{(4)}  &   \colhead{(5)}   &    \colhead{(6)}        &          \colhead{(7)} 
	}
	\startdata
	J2055-0038 & 0.05348 & 189$\rm^a$ & 40.58 & 163.67$\pm$16.01 & 216.37$\pm$35.42 & 0.60$\pm$0.20 \\ 
    J2133-0712 & 0.08659 & 167$\rm^a$ & 41.33 & -104.21$\pm$22.89 & 420.18$\pm$31.00 & 1.23$\pm$0.22 \\ 
    J2146+1113 & 0.08827 & 117$\rm^a$ & 41.48 & -42.78$\pm$25.01 & 440.44$\pm$35.45 & 2.26$\pm$0.22 \\ [0.5ex]  
    
    J0918+3439 & 0.09731 & 176$\rm^a$  & 40.80 & -398.33$\pm$2.69 & 395.11$\pm$4.87 & 1.62$\pm$0.15  \bigstrut[t] \\ 
    J1135+5657 & 0.05149 & 152$\rm^a$ & 41.62 & -172.60$\pm$1.22 & 339.79$\pm$2.56 & 1.39$\pm$0.21 \\ 
    J1404+5323 & 0.08112 & 153$\rm^a$ & 41.41 & -256.14$\pm$9.18 & 532.12$\pm$37.37 & 2.30$\pm$0.39 \\ 
    J1606+2755 & 0.04598 & 141$\rm^a$  & 40.75 & -245.93$\pm$2.00 & 296.06$\pm$2.58 & 1.16$\pm$0.09 \\ 
    J1622+3956 & 0.06303 & 117$\rm^a$& 41.46 & -15.51$\pm$2.11 & 515.75$\pm$2.58 & 1.74$\pm$0.36 \\ 
    J1720+2941 & 0.09919 & 143$\rm^a$& 41.19 & -53.14$\pm$2.11 & 414.10$\pm$3.78 & 1.92$\pm$0.37 \\ [0.5ex]  
    \tableline 
    
    J0945+1737 & 0.12829 & 113$\rm^d$ & 42.77 & -104.14$\pm$0.23 & 382.70$\pm$0.44 & 3.85$\pm$1.06  \bigstrut[t] \\ 
    J0958+1439 & 0.10912 & 242$\rm^c$& 42.60 & 0.39$\pm$0.36 & 367.82$\pm$0.67 & 3.62$\pm$0.40 \\ 
    J1000+1242 & 0.14813 & 134$\rm^c$ & 42.80 & -103.15$\pm$0.25 & 307.87$\pm$0.27 & 7.45$\pm$0.54 \\ 
    J1010+0612 & 0.09860 & 164$\rm^b$ & 42.24 & -97.28$\pm$0.53 & 549.62$\pm$0.78 & 2.36$\pm$0.44 \\ 
    J1010+1413 & 0.19944 & 306$\rm^d$ & 43.16 & -139.40$\pm$0.52 & 601.18$\pm$0.50 & 5.34$\pm$0.66 \\ 
    J1100+0846 & 0.10038 & 121$\rm^b$  & 42.78 & -4.74$\pm$0.21 & 472.05$\pm$0.38 & 3.57$\pm$0.63 \\ 
    J1125+1239 & 0.16705 & 200$\rm^c$ & 41.89 & -197.94$\pm$1.25 & 494.06$\pm$1.69 & 2.51$\pm$0.35 \\ 
    J1130+1301 & 0.13530 & 225$\rm^c$ & 41.60 & 71.54$\pm$0.52 & 316.12$\pm$0.60 & 1.20$\pm$0.31 \\ 
    J1216+1417 & 0.08176 & 85$\rm^b$ & 41.82 & 16.78$\pm$1.21 & 545.07$\pm$2.15 & 2.46$\pm$0.38 \\ 
    J1338+1503 & 0.18538 & 127$\rm^d$ & 42.60 & 168.41$\pm$0.42 & 351.47$\pm$0.84 & 5.16$\pm$0.90 \\ 
    J1339+1425 & 0.13927 & 101$\rm^b$ & 42.01 & 2.31$\pm$0.70 & 267.73$\pm$0.88 & 2.97$\pm$0.69 \\ 
    J1355+1300 & 0.15228 & 114$\rm^c$ & 41.90 & -137.00$\pm$0.79 & 351.04$\pm$0.90 & 2.91$\pm$0.76 \\ 
    J1430+1339 & 0.08518 & 189$\rm^d$& 42.65 & -23.08$\pm$0.18 & 362.12$\pm$0.32 & 3.01$\pm$0.74 \\ 
    J1504+0151 & 0.18259 & 165$\rm^c$& 42.06 & -147.48$\pm$1.24 & 476.85$\pm$1.87 & 2.52$\pm$0.65
	\enddata
	\tablecomments{(1) Name; (2) systemic redshift measured from the GMOS data; (3) estimated stellar velocity dispersion; (4) extinction uncorrected \oiii\ luminosity measured within the outflow size; (5) flux-weighted velocity shift of \oiii\ measured inside the outflow region; (6) flux weighted velocity dispersion of \oiii\ inside the outflow region; (7) kinematically measured size of outflow region after subtracting the seeing size. Note that the top 9 galaxies are observed with our GMOS programs over 2015A and 2015B semesters and the bottom 14 galaxies are selected from the GMOS archive. \\
	{$\rm^a$ based on the spatially-resolved flux-weighted stellar velocity dispersion measured at each spaxel within 3\arcsec\ aperture.}\\
	{$\rm^b$ based on the integrated spectra within 3\arcsec\ aperture.}\\
	{$\rm^c$ based on the velocity dispersion of the narrow component of \oiii.}\\
	{$\rm^d$ based on the velocity dispersion of the narrow component of \hb.}	 
	}
\end{deluxetable*}

\section{Outflow kinematics $\&$ Kinematic size}
\label{sec:outflow}

In this section, we focus on the size of ionized gas outflows based on the emission-line kinematics. To quantify the effect of outflows, we use \oiii\ velocity dispersion normalized by stellar velocity dispersion, as we used it to represent the relative strength of AGN-driven outflows in our previous studies \citep{Woo+16, Karouzos+16a, Woo+17, Kang+17}.

\subsection{Radial trend of outflow kinematics}

We present the 2-dimensional distribution and the radial profile of the normalized \oiii\ velocity dispersion, respectively, for the 9 AGNs from our observations in Figure~\ref{fig:outflow1}, and the 14 AGNs from the Gemini archive in \ref{fig:outflow2}. 
While some AGNs show a relatively symmetric distribution of \OIII\ velocity distribution, reflecting the \OIII\ flux distribution, other AGNs show strong outflows at certain directions (e.g., J0945+1737, J0958+1439, J1010+0612). In particular, J1100+0846 and J1010+1413 present strong outflows in one direction, which is consistent with the model prediction based on the biconical outflows combined with a dust plane obscuring one side of the bicone~\citep{Crenshaw+10, Bae&Woo16}. 

To investigate the radial change of the outflow kinematics, we calculate the error-weighted mean (filled blue circles), using the measurements from each spaxel (light-blue circles in Figure~\ref{fig:outflow1} and \ref{fig:outflow2}) as a function of the radial distance. Radial profiles show that gas velocity dispersion has a maximum value in the central region and gradually decreases outwards, until it becomes comparable to stellar velocity dispersion. Several AGNs show that  the central \oiii\ velocity dispersion is larger than stellar velocity dispersion by more than a factor of 4 (e.g. J1720, J0945+1737, J1100+0846, J1216+1417), while one AGN (i.e., J2055) shows no significant difference between gas and stellar velocity dispersions, indicating no or weak outflows.
    
The radial trend can be divided into two types: AGNs showing a gradual decrease of \oiii\ velocity dispersion (e.g., J0945+1737, J1100+0846, J1339+1425), and AGNs maintaining an initial plateau near the center \citep[e.g., J1010+1413, J2146, J1622; see the discussion by][]{Karouzos+16a}. To check whether this trend is due to the seeing effect, we indicate the seeing size with a vertical black dashed line in Figure~\ref{fig:outflow1} and \ref{fig:outflow2}. Approximately, a half of the sample (11 out of 23 AGNs) show that the size of the initial plateau is much larger than the seeing size, indicating that the decrease of the \oiii\ velocity dispersion occurs at a large radial distance \citep[see also][]{Karouzos+16a}.
It remains unclear why the other 12 AGNs show no initial plateau. To investigate whether the lack of the plateau is due to a relatively poor spatial resolution, we compare the spatial resolution with the outflow size, by calculating the ratio between 
the half-width-at-half-maximum (HWHM) of seeing and the outflow size (i.e., HWHM/R$_{out}$). The mean ratio is 
0.28 for the 12 AGNs without an initial plateau, while it is 0.23 for 11 AGNs with an initial plateau. The difference of the mean ratio is too small to clearly demonstrate that the non-detection of the initial plateau is due to a poor spatial resolution with respect to the outflow sizes, although we cannot rule out the seeing effect as the origin of non-detection of the plateau in individual objects. 
	
We kinematically quantify the size of outflows, using the fact \oiii\ velocity dispersion radially decreases and becomes comparable to stellar velocity dispersion, which represents the gravitational potential of the host galaxies. Although the large velocity dispersion of \oiii\ does not guarantee outflows, while it clearly represents turbulent motion, we interpret the large velocity dispersion as outflows and identify the edge of outflows, where the \oiii-to-stellar velocity dispersion ratio becomes unity. The kinematically measured outflow size by this method ranges from 0.60 to 7.45~kpc. To quantify the uncertainty of the outflow size, we adopt the range of the outflow boundary (vertical pink area) as 1$\sigma$ uncertainty, by considering the uncertainty of stellar velocity dispersion (green area in Figure~\ref{fig:outflow1}). For two AGNs, (i.e., J1606 and J1125+1239), \oiii\ velocity dispersion remains higher than stellar velocity dispersion until the edge of the photoionization region. Thus, we choose the distance of the last radial point (filled blue circle in Figure~\ref{fig:outflow2}) as the outflow size, meaning that the quoted outflow size for these object is a lower limit. 

In comparing with \OIII\ velocity dispersion, we used the flux-weighted stellar velocity dispersion measured from the spectra extracted with a 3" diameter aperture (see Section~\ref{subsec:method2} and Table~\ref{table:prop}). However, stellar velocity dispersion is expected to have a radial profile, showing a large velocity dispersion at the center as observed in nearby AGN and non-AGN galaxies with spatially-resolved kinematics \cite[e.g.][]{Kang+13, Woo+13}. Thus, to rule out the scenario that the radial decrease of \oiii\ velocity dispersion simply reflects the host galaxy gravitational potential, we compare the spatially resolved stellar and gas velocity dispersions for our 9 AGNs, by measuring stellar velocity dispersion in each spaxel and averaging them as a function of radius (blue points in Figure~\ref{fig:svd}). Note that we are not able to measure stellar velocity dispersion for the other 14 AGNs selected from the Gemini archive due to a much weaker stellar component. We find that the radial profile of \oiii\ velocity dispersion is clearly different from that of stellar velocity dispersion as stellar velocity dispersion do not steeply decrease compared to \oiii\ velocity dispersion. 
Albeit with a large scatter of stellar velocity dispersion, the maximum value of the averaged stellar velocity dispersion at the center does not exceed 300~\kms\ as expected from the range of the measured stellar velocity dispersion in AGN host galaxies \citep{Woo+04, Woo+05}, while \oiii\ velocity dispersion increases up to $\sim$500~\kms. These results demonstrate that using a constant stellar velocity dispersion measured from the integrated spectrum introduces no significant effect on determining the outflow size.

\subsection{Correlation between size and emission-line properties}

     \begin{figure}
        \center
        \includegraphics[width=0.46\textwidth]{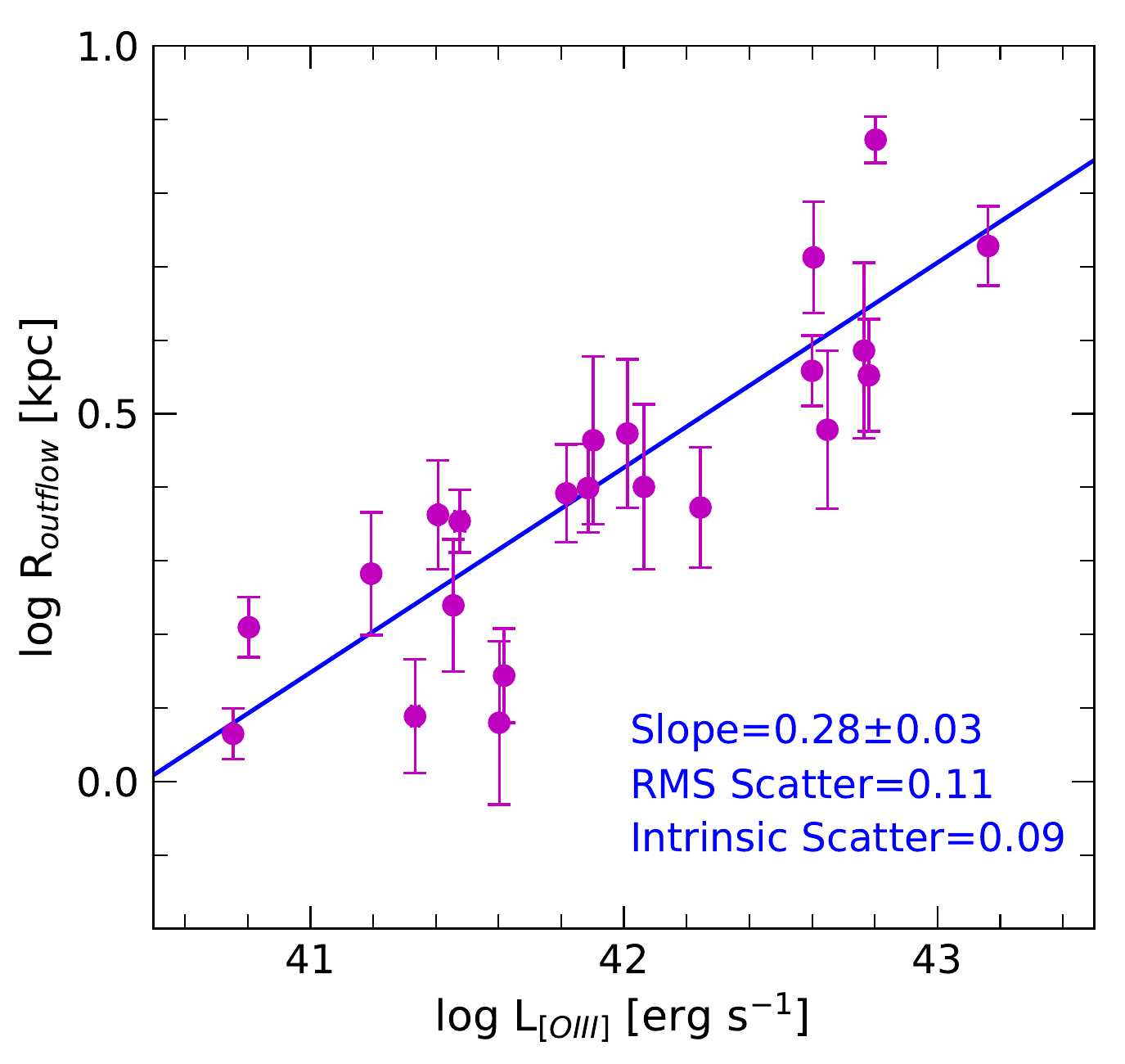}
        \caption{Correlation between the kinematically measured outflow size and \oiii\ luminosity measured inside the outflow region. The best-fit is denoted with a blue line. \label{fig:slum} }
    \end{figure}

We compare the kinematically measured outflow size with \oiii\ emission-line properties. We correct for the seeing effect by subtracting the seeing size (i.e., HWHM) from the outflow size (i.e., radius) in quadrature (Table~\ref{table:prop}), although the correction is only 4\% on average. Note that in Figure~\ref{fig:outflow1} and \ref{fig:outflow2} we presented the outflow size before the correction. We exclude one AGN (i.e, J2055), of which the outflow size is uncertain due to a lack of the radial trend of \oiii\ velocity dispersion. Two AGNs, namely, J1355+1300, J1430+1339 show distinct kinematic characteristics represented by significantly blue-shifted small separate component in the \oiii\ line profile (\citealt{Harrison+14}, see \citealt{Harrison+15} for detail). Since the results in this section are consistent with/without these AGNs, we use a total of 22 AGNs including those 2 AGNs. 

In Figure~\ref{fig:slum}, we compare the kinematically measured outflow size with the \oiii\ luminosity measured within the outflow size (see Table~\ref{table:prop}). Note that the \oiii\ luminosity integrated inside the outflow region is smaller by an average factor of 0.88 than the \oiii\ luminosity measured in the full FOV. We find a clear correlation as the outflow size increases by $\sim$0.8 dex over the 3 orders of magnitude in \oiii\ luminosity. We perform a forward regression using the MPFITEXY routine~\citep{Williams+10}, obtaining the best-fit relation:
	\begin{eqnarray}
	\rm log R_{out} &=(0.28\pm0.03) \times \rm log L_{\oiii} - (11.27\pm1.46).
	\label{eq:size_lum}
	\end{eqnarray}
We also investigate whether there exists any correlation between the outflow size and the kinematic properties of \oiii. 
Comparing the outflow size and the \oiii\ velocity dispersion measured from the integrated spectrum within the outflow size, we find no clear correlation. This result is different from that of \cite{Greene+11}, who reported a correlation between the size of the NLR (i.e., photoionization size) and the FWHM of the \oiii. 

As \oiii\ velocity dispersion represents both the gravitational component, as manifested by stellar velocity dispersion, and the non-gravitational component, we separate the outflow component using the following equation:
    \begin{eqnarray}
        (\sigma_{\oiii})^{2}&=(\sigma_{*})^{2}+(\sigma_{out})^{2}.
         \label{eq:disp}
    \end{eqnarray}
Thus, the strength of AGN-driven outflows can be represented by either the outflow velocity dispersion ($\sigma_{out}$) or the normalized \oiii\ velocity dispersion by stellar velocity dispersion ($\sigma_{\oiii}$/$\sigma_{*}$). In comparing the outflow size with the outflow velocity dispersion or the normalized \oiii\ velocity dispersion, we find no correlation. 

 \begin{figure}
    \center
    \includegraphics[width=0.48\textwidth]{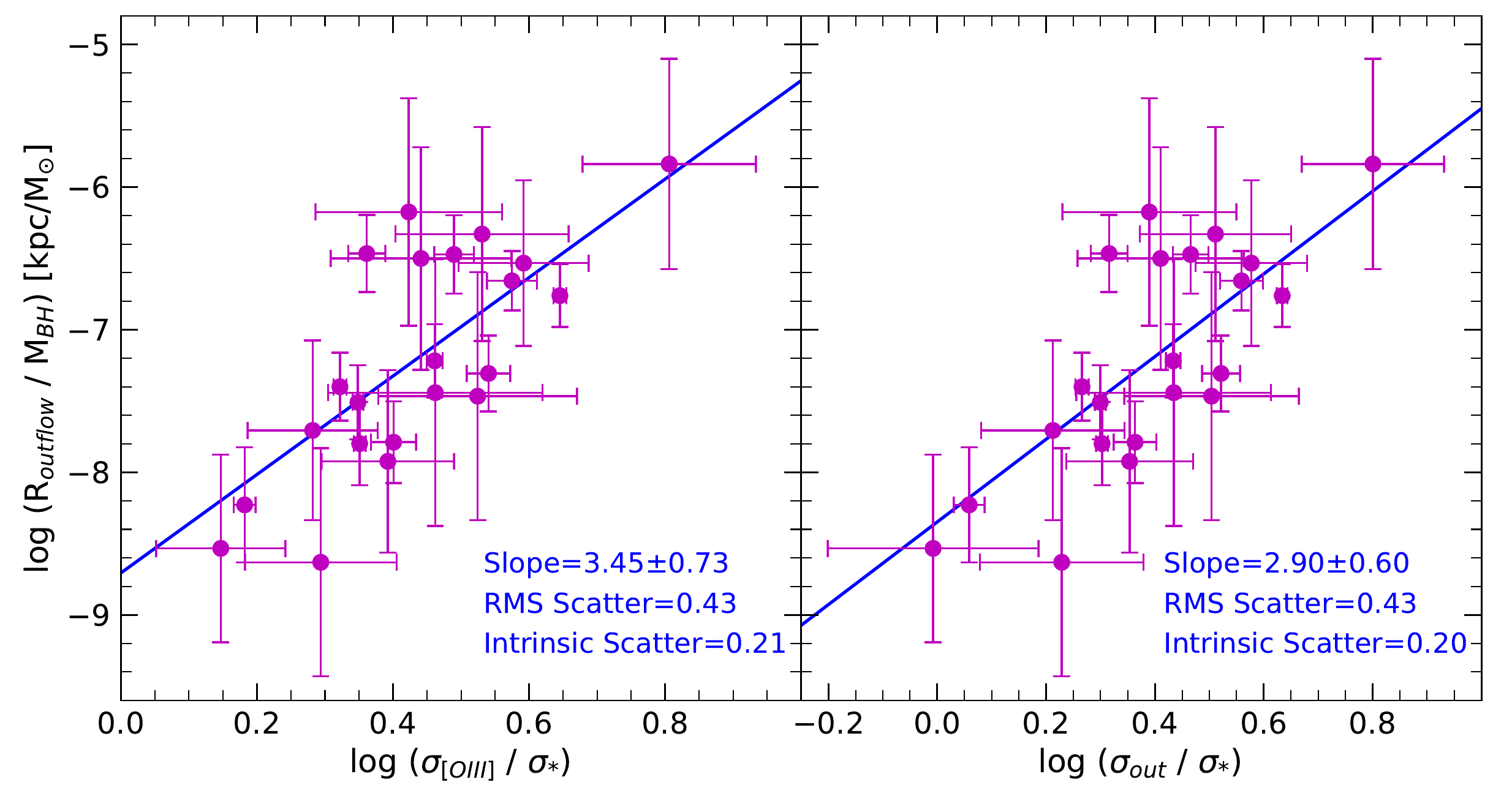}
    \caption{Comparing the outflow size divided by black hole mass with \oiii\ velocity dispersion (left) or outflow velocity dispersion (right), after normalizing them by stellar velocity dispersion. Blue solid lines represent the best fit.  \label{fig:sdisp} }
\end{figure}
 
Instead of using the measured outflow size, we now use the outflow size divided by black hole mass and compare it, respectively, with the \oiii\ velocity dispersion or the outflow velocity dispersion after normalizing them with stellar velocity dispersion in Figure~\ref{fig:sdisp}. To calculate black hole mass, we adopted the black hole mass -stellar velocity dispersion (M$_{\rm BH}$-$\sigma_{*}$) relation from \cite{McConnell&Ma13}. Note that adopting other M$_{\rm BH}$-$\sigma_{*}$ relation from the literature does not significantly change the following results. In contrast with the previous cases, the outflow size divided by the black hole mass correlates with the normalized \oiii\ velocity dispersion with the best fit slope of 3.45$\pm$0.73, albeit with large scatter.  We find a similar correlation when we use the normalized outflow velocity dispersion with the best-fit slope of 2.90$\pm$0.60, which is consistent within the 1$\sigma$ uncertainty (right panel in Figure~\ref{fig:sdisp}), as expected from the fact that outflow velocity dispersions for most galaxies are comparable to \oiii\ velocity dispersions.
Considering the correlation between black hole mass and stellar velocity dispersion, it seems that the relative outflow size for given black hole mass correlates with the non-gravitational (outflow) velocity dispersion. 

In addition, we compare the kinematic size divided by the black hole mass with Eddington ratio. For this, we calculate AGN bolometric luminosity by multiplying the extinction uncorrected \oiii\ luminosity by the bolometric correction 3500~\citep{Heckman+04}. We find a correlation between them with the best fit slope of 0.70$\pm$0.12, showing an increase of the outflow size-to-black hole mass ratio over the 3 orders of magnitude in Eddington ratio. This result suggests that more energetic AGNs have larger outflow sizes, for given the black hole's gravitational potential.

\section{Discussion}
\label{sec:discussion}

\subsection{Outflow size vs. photoionization size}
        
          \begin{figure}
            \center
            \includegraphics[width=0.42\textwidth]{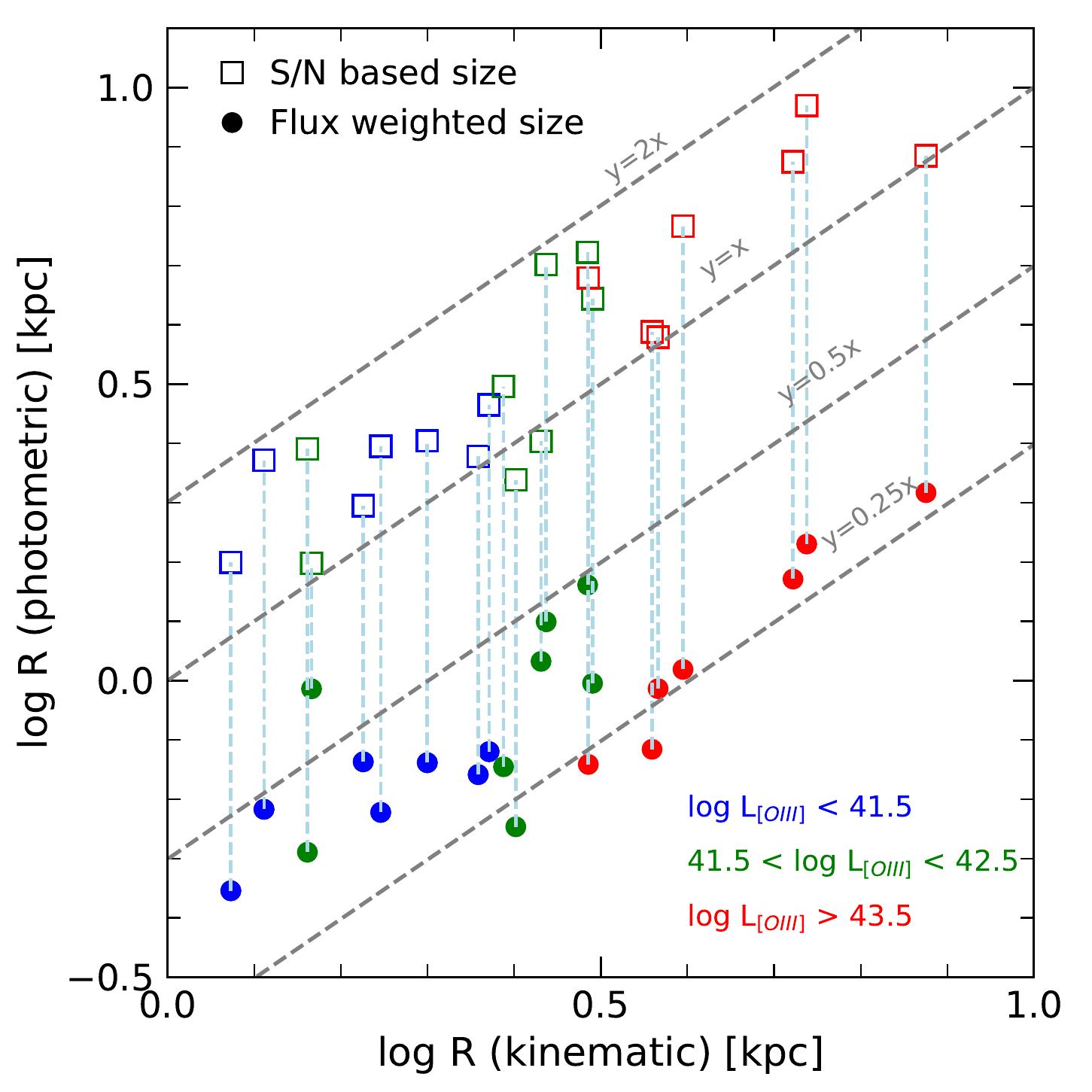}
            \caption{Comparison of the outflow size with the size of the photoionization. For given objects, two different photoionization sizes are presented: the effective radius (i.e., flux-weighted size R$_{eff}$; filled circles), and the maximum photoionization size based on the \oiii\ S/N ratio (i.e., S/N > 5) (open squares). Colors indicate three different \oiii\ luminosity bins. Each dashed line indicates one-to-two ratio, one-to-one ratio, two-to-one ratio, and four-to-one ratio between the outflow size and the photoionization size, respectively, from top to bottom.   \label{fig:scomp} }
        \end{figure}
        
        For understanding the impact of AGN feedback on galaxy scales, it is important to properly measure the size of AGN-driven outflows since the size is required for calculating the mass outflow rate and the timescale of outflows \citep{Karouzos+16b, Woo+17}. The size of the photoionized region can be measured based on the distribution of the emission-line flux (i.e., the size of the NLR). However, the size of the photoionized region is not necessarily same as the size of the outflow region since UV photons can go further out and ionize ISM beyond the boundary of the outflow region, in which outflows are dominant over the gravitational virial motion \citep{Karouzos+16a}. A recent study based on long-slit spectroscopy by \citet{Fischer+18} reported a similar conclusion that the radial size of the region, where \oiii\ shows large FWHM velocities, is much smaller than the  maximum size of the photoionized region. In contrast, in most of the previous studies, the size of the photoionized region was often adopted as the size of outflows. In this study, we properly define the outflow size by kinematically determining the edge of outflows, where outflows slow down and become comparable to stellar velocity dispersion, based on the spatially-resolved radial distribution of the \oiii\  kinematics. 
       
The \oiii\ flux maps presented in Section~\ref{subsec:emssion line flux} showed that outflow feature is detected in a smaller region than the scale where the \oiii\ is detected. The BPT maps in Figure~\ref{fig:bpt} indicated that the outflow region typically encloses the Seyfert-like photoionization region, or even extends to the LINER region. Some AGNs showed that the outflow region ends where the composite (i.e., AGN + star formation) region starts \citep[see][]{Karouzos+16b}. The difference between the dynamical timescale of outflows and the photoionization timescale investigated by \cite{Bae+17} also suggests that it is of importance to distinguish the outflow region from the photoionized region.
      
        We investigate the difference between the kinematically measured outflow size and the photoionization size using the definition reported in several studies \citep{Bennert+02, Schmitt+03, Greene+11, Husemann+14}. First, we calculate the flux-weighted effective radius (R$_{\rm eff}$) in the same manner as \cite{Husemann+14} and \cite{Bae+17} performed. Secondly, we determine the maximum (detectable) photoionization size based on the \oiii\ flux distribution, using the \oiii\ S/N ratio  $>$ 5. Note that we do not correct for the seeing size in this case, since the flux weighted effective radius is comparable to the seeing size. Figure~\ref{fig:scomp} shows that the maximum photoionization size is much larger than the outflow size, indicating that outflow size will be severely overestimated if the photoionization size is used to represent the size of outflows. On the other hand, the effective radius, which is again based on the flux distribution, is often smaller than the outflow size since the \oiii\ flux is mainly concentrated at the center. Consequently, the outflow size will be significantly underestimated, if the effective radius is used.
        
\subsection{Scaling relations with kinematic size}

        Several previous studies investigated the relation between the emission-line luminosity and the size of the NLR, which was mainly measured based on the flux distribution of \oiii. For example, \cite{Husemann+14} investigated the correlation between the flux-weighted effective radius measured from the \oiii\ flux distribution and \oiii\ luminosity based on the IFU data of 19 type 1 AGNs, reporting the best-fit slope of 0.44$\pm$0.06. \cite{Bae+17} also measured the effective radius based on the IFU data of 20 type 2 AGNs, of which \oiii\ luminosity is 1-2 order of magnitude lower than that of \cite{Husemann+14}, reporting the best slope of 0.41$\pm$0.02, which is similar to the result of \cite{Husemann+14} within the 1$\sigma$ uncertainty. In earlier studies, \cite{Bennert+02} and \cite{Schmitt+03} used the HST narrow-band images of \oiii\ and obtained the best fit slope of 0.52$\pm$0.06 and 0.33$\pm$0.04, respectively, while \cite{Greene+11} reported the slope of 0.22$\pm$0.04 from the long-slit observations of 15 radio quiet obscured quasars. The results of  \cite{Bennert+02}, \cite{Schmitt+03}, and \cite{Greene+11} strongly depend on the sensitivity of the observations because the size is defined based on the detection of \oiii\  flux, which should be larger than a certain flux limit.

The various slopes between the size of the NLR and \oiii\ luminosity reported in the previous studies are similar or steeper than the slope presented in this paper. However, the size-luminosity relation presented in this paper is physically different from that of the previous works because we measured the outflow size based on the spatially resolved \oiii\ kinematics, instead of the distribution of photoionization. The relatively shallower slope of the outflow size-luminosity relation in our study presumably reflects the efficiency of interaction between AGN power and gas in the host galaxy. The powerful energy from AGN is delivered by photons and transferred to gas, resulting in photoionization and outflows. The efficiency determines how much (and how far) gas in the host galaxy will be photoionized and how much (and how far) the kinetic energy will be transported to ISM. The shallower slope in the outflow size-luminosity relation compared to that of the NLR size-luminosity relation implies that the outflow efficiency is lower than that of the photoionization process because not all ionized gas reveals outflow features. 
       This may be due to the larger amount of energy required to push the ionized gas to make sufficient enough outflows to be detected for given the distribution of ISM and the gravitational potential of the host galaxy. As outflows extend, the interactions between the ambient matter and the outflowing gas may increase. These interactions can prevent the outflowing gas from extending further out, while the photons with enough energy may escape out and ionize the gas in a larger scale.
       
\section{Summary and Conclusion}
\label{sec:summary and conclusion}
	We used the Gemini GMOS-IFU data for 23 type 2 AGNs at z $<$ 0.2, to investigate the \oiii\ and \ha\ kinematics, and the spatial distribution of AGN-driven outflows. We kinematically measured the outflow size from the radial decrease of \oiii\ velocity dispersion and derived the outflow size-luminosity relation. Here we summarize our main results.

	\begin{enumerate}
		\setlength{\itemsep}{1pt}  
		\setlength{\parskip}{4pt}
		
		\item We measure the outflow size based on the radial profile of the normalized \oiii\ velocity dispersion by stellar velocity dispersion. The measured outflow size ranges from 0.60 to 7.45~kpc. 
        
		\item The maximum size of the photoionized region is larger than the kinematically-measured outflow size, while the flux- weighted photoionization size is significantly smaller, suggesting that using the photoionization size as a proxy for the outflow size will lead to overestimation or underestimation of the outflow size, and introduce a large uncertainty of the mass outflow rate and the energy output rate. 
        
         \item We find a correlation between the outflow size and \oiii\ luminosity with the best fit slope of 0.28$\pm$0.03, which is smaller than that of the NLR size-luminosity relation reported in the literature, which may reflect the difference of the efficiency between kinetic energy transport and the photoionization process in the host galaxies.
				
	\end{enumerate}

	Based on these results, we conclude that the kinematic size is different from the size of photoionization region, which results in different scaling relation. Therefore it is reasonable to utilize kinematic size to study ionized gas outflow in terms of probing the AGN feedback mechanism. 

\acknowledgments 
We thank the anonymous referee for the helpful comments and suggestions.
This work was supported by the National Research Foundation of Korea (NRF) to the Center of Galaxy Evolution Research (No. 2016R1A2B3011457) and POSCO Science Fellowship of POSCO TJ Park Foundation. 


\bibliographystyle{apj}




\end{document}